\documentclass[sigconf]{acmart}

\usepackage{algorithm, algorithmic}
\usepackage{graphicx}
\usepackage{subfigure}
\usepackage{arydshln}
\usepackage{xurl}
\usepackage{xcolor}
\usepackage{amsmath}

\usepackage{CJKutf8}
\usepackage{multirow}
\usepackage{url}
\usepackage{amsmath}

\newcounter{KNNumberOfComments}
\stepcounter{KNNumberOfComments}

\newcounter{JSNumberOfComments}
\stepcounter{JSNumberOfComments}

\AtBeginDocument{%
 \providecommand\BibTeX{{%
 \normalfont B\kern-0.5em{\scshape i\kern-0.25em b}\kern-0.8em\TeX}}}

\setcopyright{acmcopyright}

\copyrightyear{2023}
\acmYear{2023}
\setcopyright{acmlicensed}\acmConference[WebSci '23]{15th ACM Web Science Conference 2023}{April 30-May 1, 2023}{Austin, TX, USA}
\acmBooktitle{15th ACM Web Science Conference 2023 (WebSci '23), April 30-May 1, 2023, Evanston, TX, USA}
\acmPrice{15.00}
\acmDOI{10.1145/3578503.3583594}
\acmISBN{979-8-4007-0089-7/23/04}

\begin{document}

\title{Political Honeymoon Effect on Social Media: Characterizing Social Media Reaction to the Changes of Prime Minister in Japan}
\renewcommand{\shorttitle}{Political Honeymoon effect on Social Media}


\author{Kunihiro Miyazaki}
\affiliation{%
 \institution{Indiana University Bloomington}
 \city{Indiana}
 \country{USA}}
\email{kunihirom@acm.org}

\author{Taichi Murayama}
\affiliation{%
 \institution{SANKEN, Osaka University}
 \city{Osaka}
 \country{Japan}}
\email{taichi@sanken.osaka-u.ac.jp}

\author{Akira Matsui}
\affiliation{%
 \institution{Yokohama National University}
 \city{Kanagawa}
 \country{Japan}}
\email{matsui-akira-zr@ynu.ac.jp}

\author{Masaru Nishikawa}
\affiliation{%
 \institution{Tsuda University}
 \city{Tokyo}
 \country{Japan}}
\email{nishikaw@tsuda.ac.jp}

\author{Takayuki Uchiba}
\affiliation{%
 \institution{Sugakubunka}
 \city{Tokyo}
 \country{Japan}}
\email{takayuki.uchiba@sugakubunka.com}


\author{Haewoon Kwak}
\affiliation{%
 \institution{Indiana University Bloomington}
 \city{Indiana}
 \country{USA}}
\email{haewoon@acm.org}

\author{Jisun An}
\affiliation{%
 \institution{Indiana University Bloomington}
 \city{Indiana}
 \country{USA}}
\email{jisun.an@acm.org}

\renewcommand{\shortauthors}{Miyazaki, et al.}

\begin{abstract}
New leaders in democratic countries typically enjoy high approval ratings immediately after taking office.
This phenomenon is called the honeymoon effect and is regarded as a significant political phenomenon; however, its mechanism remains underexplored.
Therefore, this study examines how social media users respond to changes in political leadership in order to better understand the honeymoon effect in politics.
In particular, we constructed a 15-year Twitter dataset on eight change timings of Japanese prime ministers consisting of 6.6M tweets and analyzed them in terms of sentiments, topics, and users.
We found that, while not always, social media tend to show a honeymoon effect at the change timings of prime minister.
The study also revealed that sentiment about prime ministers differed by topic, indicating that public expectations vary from one prime minister to another.
Furthermore, the user base was largely replaced before and after the change in the prime minister, and their sentiment was also significantly different.
The implications of this study would be beneficial for administrative management.
\end{abstract}



\begin{teaserfigure}
\centering
 \includegraphics[width=0.90\textwidth]{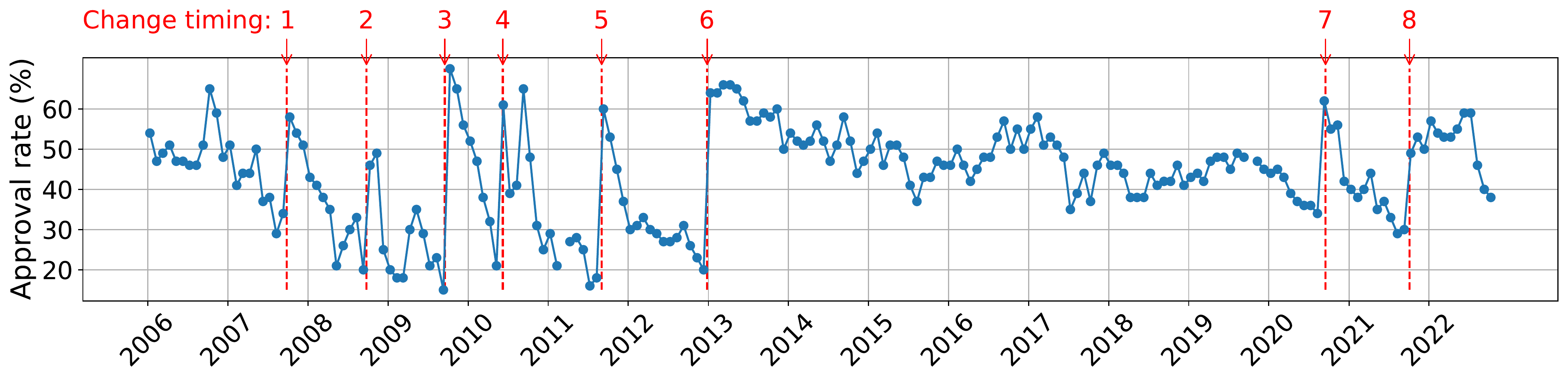}
 \caption{Monthly approval ratings for the Japanese prime ministers since 2006. The timing of the change of prime minister is annotated with a red dotted line. Data from~\cite{approvalrate}.}
 \label{fig:teaser}
\end{teaserfigure}


\maketitle

\section{Introduction}
Newly elected national leaders, such as presidents and parliamentary prime ministers (PMs), customarily enjoy high approval ratings immediately after their inaugurations.
This phenomenon, known as the \emph{honeymoon effect}, has been observed in democratic countries around the world~\cite{segatti2015renzi,qadan2022presidential,masuyama2007survival,begley2021honeymoon}.
Figure~\ref{fig:teaser} is a visualization of monthly approval ratings of the Japanese cabinet from 2006 with the timings of the PMs' change annotated, which indicates a significant jump in approval ratings at each change timing. 

Since the honeymoon effect is an important phenomenon from the perspective of government management, politicians and the news media pay great attention to it.
For example, new leaders tend to take advantage of this period of high approval ratings (i.e., the honeymoon period) as it is the best time to implement the most audacious (but potentially unpopular) policies that can leave their mark on the future economy and society~\cite{dominguez2005honeymoon,nyblade201121st,qadan2022presidential}.
Additionally, when the approval rating of the current leader declines, the administrative party attempts to preserve the people's support for the party by putting pressure on the current leader to step down, hoping for the next honeymoon effect~\cite{nyblade201121st,uchiyama2022japanese}. 

Despite its importance, the mechanism of the honeymoon effect is underexplored.
There are many narratives for why the honeymoon effect occurs, most of which come from the news media and previous studies.
For example, it is driven by the public expectation of a new leader~\cite{segatti2015renzi}, which makes the mainstream media more gentle to him/her~\cite{hughes1995not}.
On the other hand, such positivity could appear as the public appreciate the previous leader's decision to resign~\cite{abeLDP:online}, 
or treat the resignation as the virtue of integrity~\cite{kambe2017representing}.
In a nutshell, these narratives see the honeymoon effect as a change in sentiment toward two persons---a new leader and a previous leader.
However, to what extent these narratives reflect reality has not been sufficiently verified. 

In this work, we aim to deepen our understanding of the honeymoon effect by analyzing social media reactions to changes in national leadership. 
To this aim, we first collect the tweets around the changes in the PMs of Japan and construct the dataset that records the reactions to the eight changes of PMs on social media over fifteen years.
We primarily use Japanese tweets for the analysis.
Here, we would like to stress that Japan can provide a desirable case study for analyzing honeymoon effects on social media because 1) Japan has highly frequent changes of PMs compared to other countries since 2006~\cite{ryoichi2021political} when Twitter started its service, 2) Japan maintains a high degree of national and linguistic congruence~\cite{Japanese2:online}, which is easy to detect the originating country of tweets, and moreover 3) Twitter is quite popular in Japan with approximately 60M users (almost half of its population) and roughly the same number of daily active users as U.S.~\cite{Muskcall58:online}, from which we can expect a high correlation between the opinions on Twitter and the real world.

We tackle the following research questions by leveraging the dataset:

\noindent\textbf{RQ1: How does the honeymoon effect appear on social media?}
We examine whether there is a change in sentiment at an aggregated level on Twitter before and after PMs' change.
Social media allows us to extract people's raw voices, and if we can observe the honeymoon effect on social media as well, we can get insights for government management from it.

\noindent\textbf{RQ2: What aspects contribute to the change of sentiments for a new PM?}
We analyze the sentiments by topics to see what aspects people do expect from the new PMs. 
If the honeymoon effect arises from the expectations of the new PM, understanding which topics he/she is expected to address may help him/her avoid overpaying the cost of approval ratings when implementing bold policies~\cite{nyblade201121st}.

\noindent\textbf{RQ3: Who are changing the sentiments toward the prime ministers?}
We analyze the change in sentiment toward PMs at the user level.
Knowing who changes the sentiments toward PM would be helpful for the communication strategies of new PMs. 
Also, to know whether the change of PM has successfully erased the negative image associated with the previous PM may have an impact on the actions that the governing party should take when the PM's approval rating declines again~\cite{masuyama2007survival}.

As a result, we found social media tend to have a significant jump in sentiment in PM change timings (4 out of 8 times), with one timing when sentiment significantly declined and no significant results in the remaining three times. 
The sentiment was often significantly lower for PM-specific topics than for topics common to all PMs.
Also, comparisons among PMs using common topics showed that topics with positive sentiment differed by PMs.
Finally, we revealed that the majority of users were largely replaced between those who tweeted about the previous PM before the resignation and those who tweeted about the new PM after the inauguration.
Furthermore, the sentiments of these different user groups were significantly different in many cases (7 out of 8 times). 
In addition, the sentiment of users who tweeted about the same PM after the inauguration and before the resignation was significantly different (5 out of 8 times). In many cases, the sentiment was lower before the resignation (4 out of 5 times), confirming the honeymoon effect in a different way.

Our contribution is as follows: 

\noindent\textbf{1.} To the best of our knowledge, this is the first study to examine the political honeymoon effect using social media.
By looking into the past cases of the honeymoon effect, we can gain insights into how changes in administration can improve or exacerbate the quality of governance.
Moreover, the implications from this study are expected to be helpful not only in future political communication but also in other contexts such as corporate management (i.e., CEO resignation).
    
\noindent\textbf{2.} We construct a longitudinal dataset on the change of PM in Japan.
We take full advantage of Twitter data to create a 15-year tweet dataset on eight Japanese PMs consisting of 6.6M tweets.
This dataset is publicly available~\footnote{\url{https://github.com/Mmichio/honeymoon_effect_public/}}.
    
\noindent\textbf{3.} We demonstrate various data-analysis methodologies in this study. 
We use a combination of sentiment analysis and regression discontinuity design to capture the honeymoon effect.
In addition, we propose an analytical flow to distinguish between unique and common topics for each PM in the topic analysis for multiple PMs, which allows for a cross-topic analysis of historical PMs.
We believe these methods will be helpful for future research.

\section{Background and Related Works}

\subsection{Political honeymoon effect}
The honeymoon effect, in political terms, is the high approval ratings enjoyed by a new leader immediately after taking office~\cite{APreside85:online,beckmann2007policy}.
Previous research on this political honeymoon effect has been concerned with examining the existence of the honeymoon effect itself, as well as the derived effects.
\citet{dominguez2005honeymoon} found that U.S. presidents are more successful in passing legislation during their first 100 days in office than thereafter.
However, it is said that the 45th president of the US, Donald Trump---and maybe the 46th president, Joe Biden as well--- could not get the honeymoon period, and it has been debated whether a lack of honeymoon is a new normal or not~\cite{IsTheHon76:online}.
\citet{segatti2015renzi} quantitatively analyzed the honeymoon effect in Italy that led to Renzi's victory in the European elections immediately after his administration came to power.
\citet{masuyama2007survival}, noting that a PM's approval rating gradually declines after the honeymoon effect, attempted to model the relationship between the probability of a PM's resignation and his approval rating by using survival time analysis.
\citet{dewan2005corrective} analyzed that when a PM has a minister resign in the cabinet in the UK, the PM's major decision is praised, which contributes to a recovery in approval ratings, although this is not the resignation of the PM.
\citet{qadan2022presidential} analyzed the behavior of commodity markets during the U.S. honeymoon period and showed that price volatility was lower during this period. 
Additionally, studies have shown that the honeymoon effect is associated with the consolidation of democracy: the honeymoon effect is an important research topic in the field of comparative politics as well~\cite{maeda2015honeymoon, bernhard2003honeymoon}.

In this study, for the first time to our knowledge, we analyze the factors that cause the honeymoon effect to occur from a social media perspective.
In particular, we examine the change in sentiment toward a new leader and a previous leader, which has been at the heart of previous narratives on the political honeymoon effect.
Also note that the focus of this study is on the honeymoon effect, i.e., the jump in approval ratings right after the change of PMs, and not on the honeymoon period, during which high approval ratings persist. 

\subsection{Resignations of Japanese Prime Ministers}

In a parliamentary cabinet system, which the Japanese government adopts, PMs are often replaced when they lose public support and their approval ratings become quite low~\cite{ryoichi2021political}.
In particular, Japan's PMs around 2010 were replaced at a higher frequency than the heads of government in other countries, e.g., Germany and the UK, and most of them were in office for shorter terms such as about one year~\cite{ryoichi2021political}.
This is said to be because, since the late 1990s, changes in the electoral system have made the reputation of the \textit{PM as the face of the party} more important~\cite{nyblade201121st}.
That is, even from the perspective of politicians belonging to the administrative party, the PM's approval rating has a significant impact on his or her own election results; thus, when the PM's approval rate is low, there is pressure from the party members to resign, and unpopular PMs tend to have shorter terms in office~\cite{nyblade201121st,uchiyama2022japanese}. 
We regarded this frequent turnover in Japan as valuable for an object of analysis because a change of national leaders is generally a rare event that is usually difficult to analyze quantitatively.

\section{Data Collection}

\begin{table*}[!htb]
\scalebox{0.90}{
\begin{tabular}{clclccrrr}
\hline
Change &                & Resignation &                &              & Party of & \multicolumn{3}{l}{Tweet volume} \\ \cline{7-9} 
timing & Prev PM        & declaration & New PM         & Inauguration & New PM   & Prev       & New     & Total     \\ \hline
1      & Shinzo Abe     & 2007-09-12  & Yasuo Fukuda   & 2007-09-26   & LDP      & 391        & 183     & 574       \\
2      & Yasuo Fukuda   & 2008-09-01  & Taro Aso       & 2008-09-24   & LDP      & 1,538      & 2,201   & 3,739     \\
3      & Taro Aso       & 2009-08-30  & Yukio Hatoyama & 2009-09-16   & DPJ      & 7,523      & 27,457  & 34,980    \\
4      & Yukio Hatoyama & 2010-06-02  & Naoto Kan      & 2010-06-08   & DPJ      & 187,156    & 135,290 & 322,446   \\
5      & Naoto Kan      & 2011-06-02  & Yoshihiko Noda & 2011-09-02   & DPJ      & 762,486    & 206,372 & 968,858   \\
6      & Yoshihiko Noda & 2012-12-16  & Shinzo Abe     & 2012-12-26   & LDP      & 203,418    & 442,322 & 645,740   \\
7      & Shinzo Abe     & 2020-08-28  & Yoshihide Suga & 2020-09-16   & LDP      & 1,490,975  & 662,597 & 2,153,572 \\
8      & Yoshihide Suga & 2021-09-03  & Fumio Kishida  & 2021-10-04   & LDP      & 956,954    & 503,165 & 1,460,119 \\ \hline
\end{tabular}
}
\caption{The change timings of PMs, the names of the prev/new PMs, the dates of resignation declaration and inauguration, the party of new PMs (i.e., Liberal Democratic Party (LDP) and the Democratic Party of Japan (DPJ)), and the tweet volume of each change timing regarding prev/new PMs.
}
\label{table:pms}
\end{table*}

There have been eight PM changes in Japan since 2006, and we collected user reactions for each of those eight changes.
There are two key events regarding the change of PM to evoke user reactions: the previous PM's resignation declaration and the new PM's inauguration.
In this study, the period for obtaining tweets was set from 60 days prior to the previous PM's resignation declaration to 60 days after the new PM's inauguration.
It should be noted that there is a lag between these two events, but in the analysis, we do not include the data from this lag, although we obtained the data in this lag.
The tweets were acquired for the period for each change timing by using Twitter Academic API~\cite{pfeffer2022sample}.

To retrieve tweets about all PMs involved in those timings, we set the search keywords to be formal as follows (English translation is associated with Japanese words):
\begin{CJK}{UTF8}{ipxm}
\begin{itemize}
    \item Surname + given name (i.e., full name, e.g., ``岸田文雄'' (Fumio Kishida)),
    \item Surname + ``首相'' (Prime Minister),
    \item Surname + ``総理'' (Prime Minister),
    \item Surname + ``政権'' (Administration),
    \item Surname + ``内閣'' (Cabinet).
\end{itemize}
\end{CJK}
In order to increase the precision of the detection of tweets about the PMs and cabinets, we did not include keywords of only surnames, surnames with casual titles (e.g., Kishida-san), and other nicknames.

We did not include retweets in the collection since our purpose was to capture user sentiment.
Also, since Japanese is mostly spoken in Japan~\cite{Japanese2:online}, we did not conduct additional filtering based on the country.
As a result, we collected around  6,557,452 tweets written by 855,694 unique users. 

As a post-processing of tweet acquisition, we conducted user-based removal of inappropriate tweets. 
First, since we were interested in the sentiment of general users, we excluded accounts marked as verified~\cite{edgerly2019blue}\footnote{Note that we finished our data collection before the Twitter acquisition by Elon Musk, which affects the quality of verified mark~\cite{ElonMusk68:online}.}.
Next, to preserve ordinary users' accounts, we removed the following accounts:

\begin{CJK}{UTF8}{ipxm}
\noindent\textbf{News media:}
accounts that have ``ニュース'' (news), ``news,'' ``新聞" (newspaper), and ``テレビ'' (TV) in the account names, and

\noindent\textbf{Organizations:}
accounts that have ``公式アカウント'' (official account) in the description.
\end{CJK}

We also used Botometer~\cite{davis2016botornot} to remove bot-like accounts.
Botometer provides scores from 0 to 1 to measure how the accounts are likely to be bots.
There are two types of scores: one for English and the other for non-English (the language-independent score), among which we used the latter one.
Since the prior studies suggested the threshold of Botometer score to classify bots and humans mainly targeted English texts and even the threshold of the score for English accounts varied to some extent~\cite{rauchfleisch2020false}, we decided to verify the accuracy with a bootstrap method to decide a threshold.
Specifically, for four groups with scores of (0.8-0.85], (0.85-0.9], (0.9-0.95], and (0.95-1.00], we randomly sampled 30 accounts for each group, and three Japanese authors manually annotated whether the account was considered to be a bot or not
As a result of majority voting, 13.3\% (Fleiss' Kappa score: 0.24), 26.7\% (0.19), 43.3\% (0.49), and 63.6\% (0.45) were determined to be bots, respectively.
Therefore, in order to remove obvious bots, 0.95 was set as the threshold for this study, and accounts with scores higher than that were removed. 
We note we could not remove all bots in our data, which is a limitation of this study. 
Nonetheless, the ratio of (0.95-1.00] users is 0.5\%, and one of even (0.9-0.95] users is 2.2\%, and we believe there is no significant impact on the results.
In addition, to remove low active accounts, we removed accounts with a total tweet volume of less than ten (according to their profile information) at the time of data acquisition (i.e., Nov. 2022).
At this point, the number of tweets was 6.1M, and the number of users was 838k.

Finally, to focus on the sentiment toward a single PM, tweets containing the surname of more than one PM were excluded.
This results in 5,590,028 tweets by users 818,043 users. We use this dataset for the rest of the analyses. 
Table~\ref{table:pms} summarizes the name of the previous and new PMs, the date of declaring resignation, the date of inauguration, and the number of tweets for each change timing.
We note that the administrative party changed twice in those eight change timings between the Liberal Democratic Party (LDP) and the Democratic Party of Japan (DPJ).
Also, the dataset includes Abe's two appointments.

\section{RQ1: How does the honeymoon effect appear on social media?}

In this section, we examine whether the honeymoon effect also occurs on social media.
Specifically, we analyze the change in sentiment toward the PM before and after the change of PM using sentiment analysis and a regression discontinuity design (RDD).

\subsection{Sentiment Analysis}
\noindent\textbf{Predictive Model:}
We first predict the sentiment of each tweet.
As a predictive model, we use Asari, an open-source sentiment quantification model for Japanese sentences~\cite{Hironsan72:online}.
Asari is an SVM-based model that returns a sentiment value between 0 and 1 (the closer to 1, the more positive) given a Japanese sentence. 
We chose Asari because it is reported to perform with compelling accuracy as BERT-based models and is fast~\cite{nihongoSenti59:online}.
Although dictionary-based methods such as LIWC~\cite{pennebaker2015development} (or J-LIWC for Japanese texts~\cite{igarashi2021development}) are frequently used for sentiment analysis, we did not use those methods because we found that some PMs had only a few tweets with sentiment words in the dictionary.
As a bootstrap confirmation, we randomly sampled 30 tweets each from the upper half (i.e., more than the median score) and the bottom half (i.e., less than the median score) in terms of sentiment to see whether a Japanese author could classify them, resulting in an F1 score of 0.73.

\noindent\textbf{Aggregation of sentiment scores:}
We aggregate sentiment on a daily basis because the number of tweets may vary by time of day.
Also, to mitigate the influence of the more vocal users who tweet multiple times a day, we first averaged the sentiment of each user per day, and then averaged the sentiment of all users who tweeted that day.
The sentiment $S$ for a day $t$ for a given PM $p$ is depicted as follows: 
\begin{equation} \label{eq:senti}
S(t,p) = \sum_{u:user}\frac{\sum_{d\in D_{u,p,t}}s(d)/|D_{u,p,t}|}{|U_t|}
\end{equation}
where $t$ is the time (day), $p$ is the PM, $d$ is the document (tweet), $s(d)$ is the sentiment score of the tweet $d$ by Asari, $D_{u,p,t}$ is the set of the user $u$'s tweet about $p$, and $U_t$ is the set of all users who tweeted on that day $t$.

\subsection{Regression Discontinuity Design}
To see the change in sentiment before and after the change of PM, we perform a regression discontinuity design (RDD).
We employ a linear model:
\begin{equation} 
S_t=\alpha_0+\beta_0 t+\alpha i_t+\beta i_t t+\epsilon_t
\end{equation}
where $S_t$ is the sentiment toward a given PM on a given day $t$ (eq~\ref{eq:senti}), $t$ is a date, which takes values from -60 to +60, starting 60 days before the previous PM's declaration of resignation and ending 60 days after the new PM's inauguration date, and $i_t$ is an indicator variable, which is 1 after the change of PM (i.e., $t>0$) and 0 otherwise.
Here, the inauguration date ($t=1$) is hypothetically set to be the next day of the date of the resignation declaration ($t=0$).
We assumed the lag between the actual resignation declaration and the inauguration date to be a ``grace period'' (following ~\cite{horta2021platform}), and the data for that period is omitted because PM's resignations are often sudden and tend to cause instability in sentiment. 
This model allows us to capture not only the change in the quantity of sentiment before and after the change of PM, represented by $\alpha$, but also the trend in sentiment before and after the change, represented by $\beta_{0}$ and $\beta$; thus, we can understand the change in sentiment more deeply than by simply comparing the average of sentiment before and after change timing.
Lastly, $\epsilon_t$ is an error term at $t$. 

In the introduced RDD model, we input the sentiment of the previous PM before the resignation declaration and the sentiment of the new PM after the inauguration into the model.
This is in line with the traditional survey that changes the targets from the previous PM to the new PM before and after the change of PMs.

\subsection{Results}
\begin{figure*}[!htbp]
\centering
\includegraphics[width=0.92\linewidth]{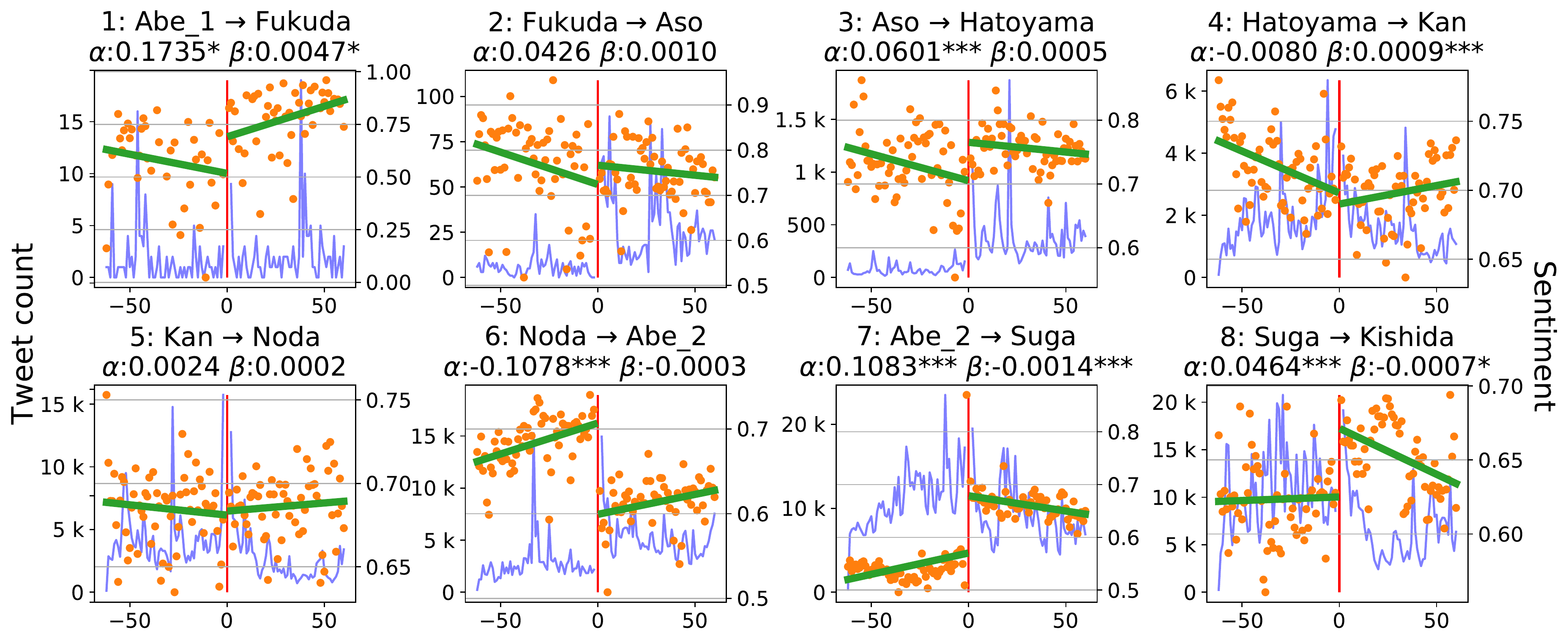}
\caption{
Change in daily sentiment before and after the change of PMs.
The left side of the red line shows the period before the PM announced the resignation, and the right side shows the period after the new PM took office.
The blue line shows the daily volume of tweets, and the orange dots show the average daily sentiment, with the left side about tweets about the former PM and the right side about tweets about the new PM.
The green line depicts the model fitted to the sentiment in the regression discontinuity design (RDD).
Above the subplots, we report coefficients associated with the PMs' change in the model ($\alpha$ and $\beta$). 
Coefficients for which $p<$0.001, 0.01, and 0.05 are marked with ***, **, and *, respectively.
}
\label{fig:RDD}
\end{figure*}

Figure~\ref{fig:RDD} shows the results of the RDD.
The $\alpha$ of each subplot shows that sentiment increased significantly in four of the eight change timings (change timing: 1,3,7,8), while three did not show a significant change (2,4,5), and one showed a significant drop (6). 
In other words, we observed that social media tends to show the honeymoon effect, but not at every change timing.
While it has been known that the demographics of Twitter are not necessarily the same as in the real world~\cite{an2015whom}, this result demonstrates a possibility of utilizing social media for the analysis of the honeymoon effect using sentiment toward the PM.

It is also interesting to see clear jumps in sentiment in the last two changes (Figure~\ref{fig:RDD}-7: Abe$\rightarrow$Suga; Figure~\ref{fig:RDD}-8: Suga$\rightarrow$Kishida). 
Abe's second administration lasted about eight years, the longest among all the Japanese PMs in history, and the number of Twitter users increased during that time. 
In fact, the number of tweets jumped dramatically during the last two change timings (Table~\ref{table:pms}).
In other words, the demographics of Twitter might have approached the actual population over the eight years, and the fact that the honeymoon effect was clearly confirmed twice in these circumstances may suggest that the correlation between changes in actual approval rating and sentiment has increased because the number of Twitter users increased.

\section{RQ2: What aspects contribute to the change of sentiments for a new PM?}

In RQ1, we found that sentiments toward PMs significantly change at multiple change timings of PMs. 
Here we examine from which aspects the sentiment causes these changes using topic modeling.

\subsection{Topic Modeling}\label{sec:topic_model}
\noindent\textbf{Model selection and data:}
We use the Biterm Topic Model (BTM)~\cite{yan2013biterm} for the topic modeling.
This model is a derivative of LDA~\cite{blei2003latent} and is known to be able to extract topics with higher accuracy for short sentences.
For this analysis, we focus on the tweets during the 60 days after the inauguration of PMs at each change timing.
We combine the corpus for all new PMs for the later comparative analysis of topics across PMs.
To reduce the imbalance in data volume, we limit the number of tweets for each PM to (randomly sampled) 30,000 for the model training.
Then, we remove the stopwords from texts and leave only nouns~\cite{martin2015more}, and use BTM to label each tweet with the topic with the highest possibility calculated by BTM.
After training the model, we predict the topic of tweets that were not used for training and include them for later analysis.

\noindent\textbf{Number of topics and separation of unique and common topics:}
To determine the number of topics, we set the number of topics large at the beginning and then merge them to find the optimal number of common topics as in~\cite{balasubramanian2022leaders}.
Here, it is possible that some topics are unique to some PMs, and some are common to all PMs.
Topics common to all PMs allow for comparison of sentiment across PMs, but this is not the case for unique topics.
Therefore, we leverage the topic number selection method to classify unique topics and common topics in the process of merging topics.
Specifically, for each topic, we calculated $R(T,p)$, the ratio of the amount of text as for topic $T$ assigned to each PM $p$. 
Then, if $R(T,p)$ of a PM $p$ about a fixed topic $T$ outstands among PMs, this topic can be considered a unique topic for PM $p$.
We calculate $R(T,p)$ while addressing the imbalance in tweet volume among PMs.
To calculate $R(T,p)$, we first compute the ratio $r(T,p)$: how dominant a topic $T$ is from the tweets for a PM $p$.
\begin{equation}
r(T, p) = \frac{\left|\left\{\text{tweet} \in T\mid \text{tweet for a PM }p\right\}\right|}{\sum_{T':\text{topic}}\left|\left\{\text{tweet} \in T'\mid \text{tweet for a PM }p\right\}\right|}.
\end{equation}
Next, we calculate $R$ by modifying $r$ so that the sum of $r$ among PMs is 1: 
\begin{equation}
R(T,p) = \frac{r(T,p)}{\sum_{p':\text{PM}}{r(T,p')}},
\end{equation}
i.e., $R(T,p)$ is the normalized version of $r(T,p)$.
Here, assuming that tweets about each PM are equally assigned in terms of a given topic $T$, the expected value of this score $R$ should be equally divided by 8 and be 0.125.
We divide the topics generated by BTM into unique and common topics using a certain threshold of max $R$ among the $R$ scores for all PMs. 

We tested 25, 50, 75, 100, and 125 topics and chose 75 as the initial number of topics by comparing the perplexity scores~\cite{zhao2015heuristic}.
Then, we conducted BTM for all the tweets within the period of 60 days after all the inaugurations.
In order to set the threshold for $R$ to classify the unique and common topics for PMs, we sorted the topics by the max value of PMs' $R$ scores.
After the first author, who is fluent in Japanese, manually examined the top representative words of each topic in terms of the possibility of belonging assigned by BTM, we decided the threshold as 0.285, below which the topics are no longer unique for any of the PMs.
In other words, for a topic $T$, the smaller the max $R$ is, the flatter the distribution of $R$ for all PMs becomes.
Finally, the Japanese authors manually checked and replaced the labels of a total of 12 topics with the $R$ scores around the threshold (7 from unique and 5 from common topics).
As a result, we obtained 52 unique topics and 23 common topics for all the PMs.
As for unique topics, we got 5 topics for Fukuda, 5 for Hatoyama, 4 for Kan, 8 for Noda, 3 for Aso, 10 for Abe, 7 for Suga, and 10 for Kishida.
The details on common topics are shown in Table~\ref{table:common_topics} in the appendices, while we omitted the details on unique topics for the reason of space. 

\noindent\textbf{Merging Common Topics:}
Some of the 23 common topics overlap in terms of their contents; thus, we merge them to find an optimal number of common topics.
Here, we propose a clustering method to systematically merge similar topics.
Specifically, we first connected ten representative words of each topic into one pseudo-sentence and made 23 sentences.
Then, we use the trained BTM to calculate the probability of each sentence belonging to each topic and calculate the Hellinger distance between the sentences using the probabilities~\cite{srivastava2009text}.
Lastly, we make the distance matrix based on the Hellinger distance of the sentences and conduct hierarchical clustering with Ward's method~\cite{murtagh2014ward} to aggregate similar sentences. 
We manually determined the threshold for aggregation by hierarchical clustering by looking at the dendrogram and the contents of resulting clusters (Figure~\ref{fig:dendrogram} shown in the Appendices). 
We note that we also tried clustering based on the average of the top word embedding by BERT model~\cite{devlin2019bert,cltohoku73:online} and Euclidean distance between topics, but the clusters made more sense when using the topic distribution by BTM and the Hellinger distance. 


As a result, we obtained four clusters. We label them as \textit{Diplomacy}, \textit{Economy}, \textit{Cabinet personnel (Personnel)}, and \textit{Diet administration (Diet)} (i.e., handling and running the government) by examining the representative words of the topics that comprise the cluster (shown in Table~\ref{table:common_topics} in the Appendices).

\subsection{Results}

First, we compare the sentiments of common and unique topics.
Expectations and criticisms of the PM can occur on both common and unique topics.
For example, it is natural to assume that the economy and diplomacy are common topics among PMs. On the other hand, occasional topics, such as disasters and epidemics, and personal topics, such as scandals, would be considered unique topics.
Depending on which sentiment is lower, common topics or unique topics, the government's policy would vary.
For example, if sentiment on common topics is low, there is much that should be learned from past PMs. 
On the other hand, when unique topics indicate low sentiments, especially scandals, the PM and the government would elaborate unique solutions for the matter.  

Table~\ref{fig:common_vs_indiv} shows the average sentiments of common/unique topics for each PM with the results of the Mann-Whitney U test.
We note that we use a user-based daily average of sentiments to mitigate the influence of extremely active users in the same way as eq~(\ref{eq:senti}).
The results show that sentiment is largely lower for unique topics: four PMs have significantly lower sentiment for their unique topics, two PMs have significantly higher sentiment for common topics, and two PMs do not have significant results.
Here, we look at the unique topics with the lowest sentiment for four PMs who had lower sentiment for unique topics: 
Kan for consumption tax hike~\cite{Japansne12:online} (sentiment value: 0.643),
Noda for nuclear power plant operation~\cite{Analysis96:online} (0.626), 
Suga for pressure on academia (i.e., the Science Council of Japan)~\cite{EDITORIA16:online} (0.640), and
Kishida for unconditional cash-only handouts for COVID-19~\cite{KishidaF51:online} (0.542).
The results actually show that sentiment was low on issues that were unique to the time or the PM's personal policy. 
Conversely, when the honeymoon effect occurs on social media, it can often be caused by a common topic. 

\begin{table}[!htbp]
\centering
\scalebox{0.88}{
\begin{tabular}{lccc}
\hline
New PM         & Common & Unique & Significance                                     \\ \hline
Fukuda   & 0.848  & 0.849  & -                                          \\
Aso      & 0.781  & 0.885  & \textless{}\textless{}\textless{}          \\
Hatoyama & 0.782  & 0.828  & \textless{}\textless{}\textless{}          \\
Kan      & 0.751  & 0.736  & \textgreater{}\textgreater{}\textgreater{} \\
Noda     & 0.738  & 0.734  & \textgreater{}                             \\
Abe\_2   & 0.636  & 0.636  & -                                          \\
Suga     & 0.721  & 0.698  & \textgreater{}\textgreater{}\textgreater{} \\
Kishida  & 0.704  & 0.679  & \textgreater{}\textgreater{}\textgreater{} \\ \hline
\end{tabular}
}
\caption{
A comparison of sentiment on common and unique topics For each PM.
Significance shows the results of the Mann-Whitney U test, where the number of brackets indicates 3: $p < 0.001$, 2: $p < 0.01$, 1: $p < 0.05$.
The direction of the brackets indicates the greater values (e.g., for Aso, unique topics have higher sentiment than common topics).
}
\label{fig:common_vs_indiv}
\end{table}

Next, we conduct a comparison of sentiments between PMs about common topics.
Here, since the average sentiment differs from topic to topic and from PM to PM, we correct the sentiment values in order to make the comparison.
In particular, we create a PM-topic sentiment matrix, calculate the expected value $E(T,p)$ for each cell in the matrix, and then calculate the deviation of each sentiment value from that expected value (i.e., residuals)~\cite{sharpe2015chi}.
If this deviation is positive, it means that the sentiment is higher than expected.
The expectation and residual can be calculated as follows:
\begin{align}
E(T,p) &= \sum_{T}{\sum_{p}{S_{T,p}}} \times \frac{\sum_{p}{S_{T,p}}}{\sum_{T}{\sum_{p}{S_{T,p}}}} \times \frac{\sum_{T}{S_{T,p}}}{\sum_{T}{\sum_{p}{S_{T,p}}}}, \\
Residual_{T,p} &= (S_{T,p} - E_{T,p}) / \sqrt{\smash[b]{E_{T,p}}}, \label{eq:residual}
\end{align}
where $S_{T,p}$ means the average of tweets' sentiment for a topic $T$ and a PM $p$. 

\begin{figure}[!htbp]
\centering
\includegraphics[width=0.87\linewidth]{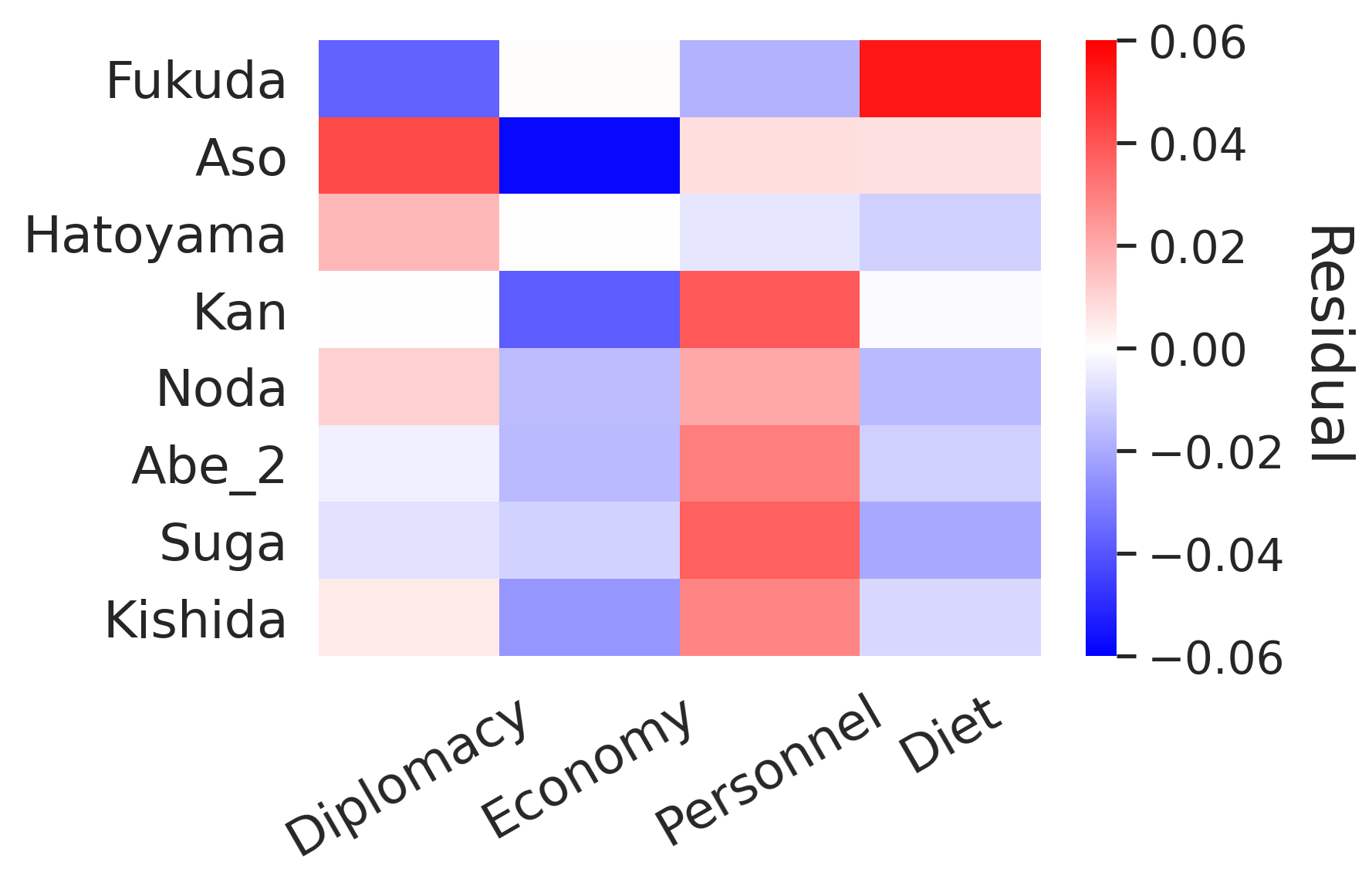}
\caption{
Average sentiments of the tweets with respect to each PM for each topic.
The color of each cell corresponds to the value of residual (eq~\ref{eq:residual}); the redder the color, the higher the sentiment. 
We clipped the values above an absolute value of 0.06 for visibility.
}
\label{fig:topic_sentiment}
\end{figure}

Figure~\ref{fig:topic_sentiment} is a heatmap visualizing the residuals, which shows that sentiment value varies by topic.
First, we could observe the overall trend in the figure: in the last four PMs, the sentiment was basically high on Personnel and low on Economy.
It makes sense that cabinet personnel is attracting higher sentiments since it has been said that the attitude of Japanese people toward politics weighs on ``personality rather than policy''~\cite{curtis2004japanese,miyazaki2021retrospective}, meaning that \textit{who} is in power is important to the Japanese people.
Also, as for economic topics, the outlook of Japanese people for the future economy is known to be highly pessimistic, especially for the next generation's economy, with 76\% of respondents saying it will get worse off, which was the worst among the 27 countries surveyed by the Pew Research Center~\cite{Japanese25:online}.


Looking at the comparison of PMs in Figure~\ref{fig:topic_sentiment}, we could confirm the differences in sentiments by topics. 
For example, Kishida was said to be expected of high diplomatic skills~\cite{WhoisFumioKishida:online}, which corresponds to the result that there is positive sentiment for him.
Conversely, Kishida has been criticized for the cash handouts for COVID-19 that he proposed at the beginning of his term~\cite{KishidaF51:online}, indicating that there are low expectations for the economy.
In addition, Suga was known to have been favorably received about the appointment of cabinet ministers~\cite{Newappoi9:online}, which corresponds to his higher sentiment score toward Personnel.
These results suggest an alignment between the actual political situation and the initial public expectations on social media, which would indicate the topics that should be anticipated first at the timing of the PM's inauguration. 


\section{RQ3: Who are changing the sentiments toward the prime ministers?}

For the administrative party, one of the intentions of the change of PM is to aim to calm users with negative feelings toward the PM.
However, the current hearing survey system for PM's approval rating does not track individual users, making it difficult to determine whether people with negative sentiment have decreased or people with positive sentiment have increased.
Therefore, in this section, we conduct an analysis to capture changes in the sentiment at a user level by tracking individual users before and after the change of PMs.

\begin{figure*}[!htbp]
\centering
\includegraphics[width=0.89\linewidth]{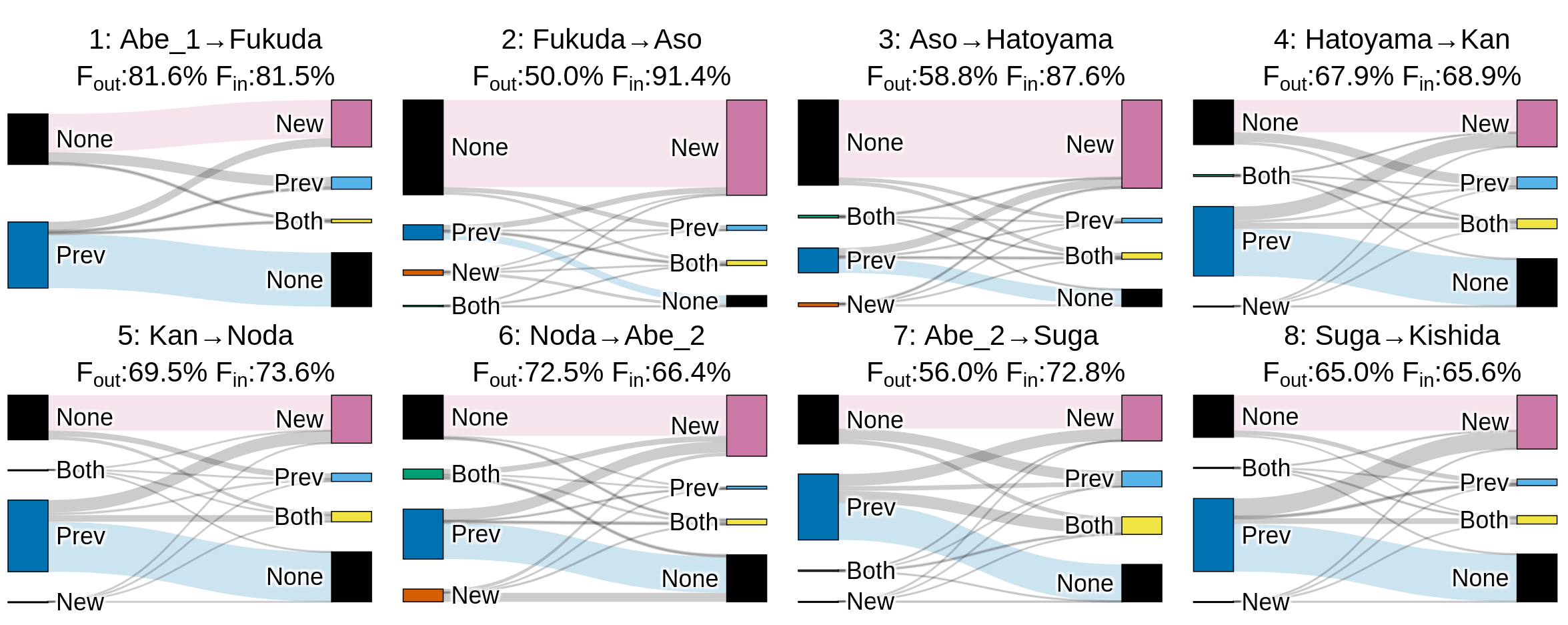}
\caption{Sankey diagrams indicating user group transitions at each change timing. 
At each Sankey diagram, each rectangle represents the size of the user base mentioning the prev/new PM.
The left side shows the users before the resignation of the previous PM and the right side shows the users after the inauguration of the new PM.
``Both'' indicates users tweeting about both PMs, ``Prev'' indicates users tweeting about the previous PM, ``New'' indicates users tweeting about the new PM, and ``None'' indicates users not tweeting about either PM.
The line connecting the rectangles shows the transition of the user base from left to right.
$F_{out}$ indicates the percentage of users tweeting neither PM among users who tweeted about the previous PM before his resignation, and $F_{in}$ indicates the percentage of users who newly started tweeting about the new PM among those who 
 did not tweet about either PM before the inauguration.
}
\label{fig:user_transition}
\end{figure*}

Here, we examine whether there is an overlap in the user base of the previous and new PMs' tweets.
Specifically, we first collected tweets for each change timing for two periods: 60 days before the resignation of the former PM (Before) and 60 days after the inauguration of the new PM (After).
Then, we classified the authors of the tweets of each change timing into four groups: \textit{Prev}, users who tweeted only about the previous PM; \textit{New}, users who tweeted only about the new PM; \textit{Both}, users who tweeted about both the previous and the new PM; \textit{None}, users who did not tweet about either the previous and the new PM.
Figure~\ref{fig:user_transition} shows the transition of these users Before and After.  

We found the two types of users dominate the majority of users in each change timing: Prev$\rightarrow$None, i.e., users who tweeted about the previous PM in Before but did not tweet about either the previous PM or the New PM in After; None$\rightarrow$New, i.e., users who did not tweet about either the PM in Before but tweeted about only the New PM in After.
To measure the presence of these users, we computed the following two indices:
\begin{align*}
F_{out} &= 100 \times \frac{|\text{Before\_Prev} \cap \text{After\_None}|}{|\text{Before\_Prev}|}\\
F_{in} &= 100 \times\frac{|\text{Before\_None} \cap \text{After\_New}|}{|\text{After\_New}|}
\end{align*}
where $\text{A\_B}$ is the user group $B$ in the period $A$.
$F_{out}$ indicates the percentage of users who stopped tweeting about either PM among those who tweeted about only the previous PM.
$F_{in}$ is the percentage of users who did not tweet about either PM among those who newly tweeted about only New PM.
These two indicators are annotated in Figure~\ref{fig:user_transition}. 

The figure shows that $F_{out}$ is mostly above 50\% and $F_{in}$ is all above 65\%, indicating that the majority of users who mention the previous and new PMs are being replaced before and after a change timing of PMs.
In other words, when there is a change in sentiment before and after a change timing of PMs, it is more likely that users with different perceptions have replaced the whole opinion rather than a change in the same individual's state of mind.

\begin{figure}[!htbp]
\centering
\includegraphics[width=0.91\linewidth]{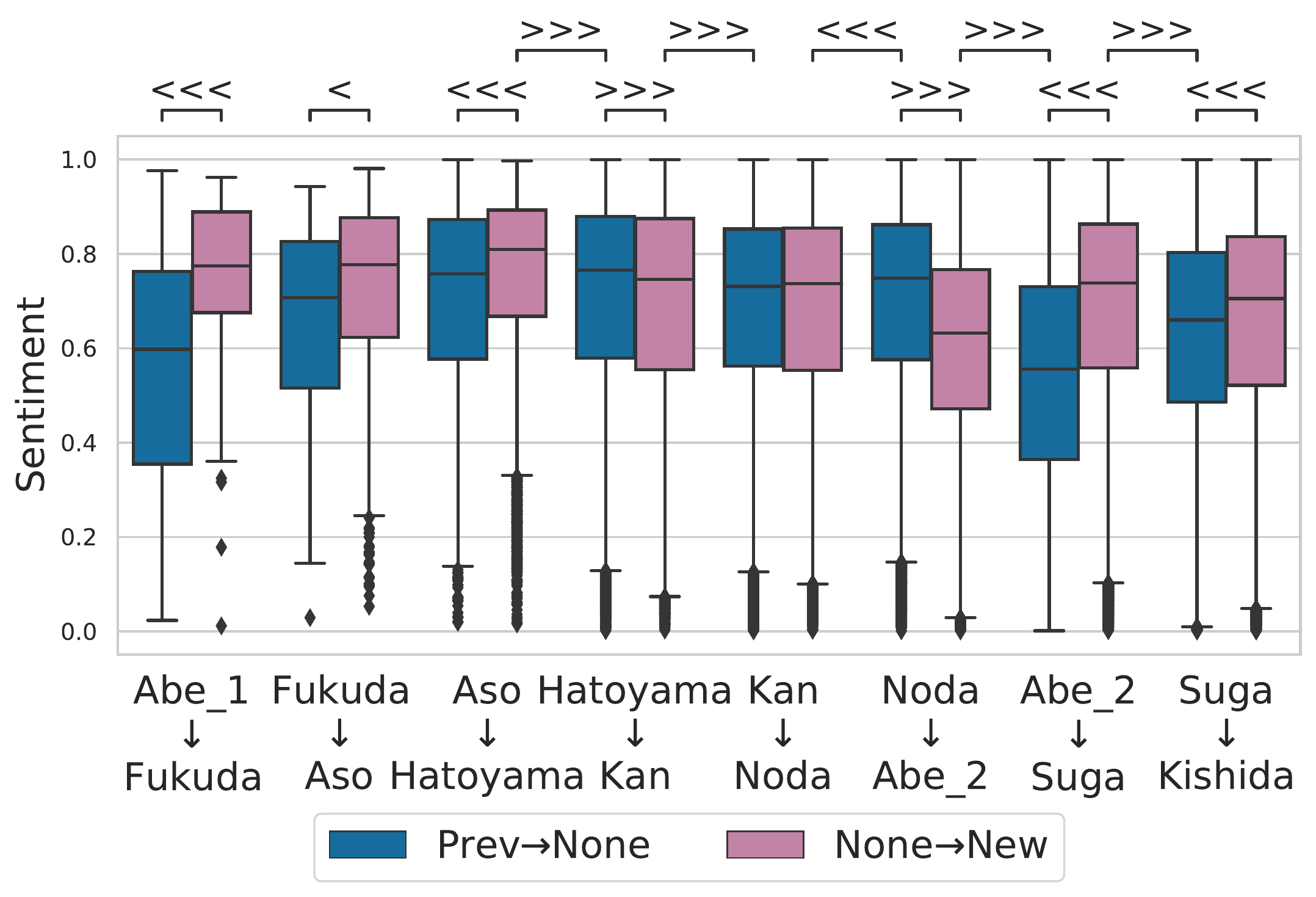}
\caption{
Boxplot of sentiment for groups of user transitions at each change timing.
The sentiment of the groups of Prev$\rightarrow$None and None$\rightarrow$New are compared in each change timing.
Also, the sentiments for the same PM in different change timings are also compared. 
The number of brackets indicates the result of the Mann-Whitney U test: 3: p < 0.001, 2: p < 0.01, 1: p < 0.05.
The direction of the brackets is greater for the direction of the open side of the brackets. 
Visualization by~\cite{florian_charlier_2022_7213391}.
}
\label{fig:boxplot_sentiment}
\end{figure}

We also compare the sentiment among groups in user transition.
We chose the groups for the comparison as the Prev$\rightarrow$None and None$\rightarrow$New groups, which are in the majority of the user transitions.
In addition, we compare the sentiments of the same PM as New PM and as Prev PM, i.e., we compare the None$\rightarrow$New groups and Prev$\rightarrow$None for the same PMs.
Figure~\ref{fig:boxplot_sentiment} shows the result of the comparison.

First, comparing Prev$\rightarrow$None and None$\rightarrow$New at the same change timing, we find that there was a significant difference in sentiment in seven out of eight times.
This means that the majority of tweets are substituted by a group with different sentiments.
In five of those timings, the None$\rightarrow$New group had higher sentiment, and in 2 timings, the None$\rightarrow$New group had lower sentiment.
This was in almost perfect agreement with the timing at which the jump in Figure~\ref{fig:RDD} was confirmed.
In other words, the change in sentiment that occurred around the time of change timings was confirmed to be largely due to the substitution of these groups of tweets. 

Furthermore, a comparison of sentiments about the same PM at different timings shows that there was a significant difference in five out of eight cases.
Moreover, four of the eight times, Prev$\rightarrow$None sentiment was significantly lower, meaning that the PM's sentiment that was high when he took office was lower just before he resigned.
This could be a case of the honeymoon effect being confirmed in another way. 

Here, we should note that only Abe exhibits extremely low sentiment.
Abe has apparently lower sentiment than other PMs at the end of his first administration, as well as at the beginning and end of his second administration.
It seems that the low sentiment at the end of the first administration carried over into the second administration.
While a PM's name recognition often jumps when they become PM, Abe's name recognition was already strong at the beginning of his second administration. 
This result may explain why the RDD results in Figure~\ref{fig:RDD} indicate a drop in sentiment only when Abe started his administration. 

Finally, we compare the same PM at the same change timing.
Here, we compare Prev$\rightarrow$None and None$\rightarrow$Prev for the same PM.
The PMs resign as a form of taking responsibility for their low approval ratings, and we test whether this eases the harsh atmosphere against them.  
Table~\ref{fig:prev_new} shows the results of the comparison.
We got the result that three out of eight times, the sentiment of None$\rightarrow$Prev is significantly higher than that of Prev$\rightarrow$None (Hatoyama, Abe\_2, and Suga).
Conversely, on one timing, None$\rightarrow$Prev was lower (Kan).
This indicates that there is a tendency for some users to emerge to appreciate a previous PM's labor, although not always.


\begin{table}[!htbp]
\begin{tabular}{lccc}
\hline
Prev PM  & Prev=\textgreater{}None & None=\textgreater{}Prev & Significance                               \\ \hline
Abe\_1   & 0.561                   & 0.609                   & -                                          \\
Fukuda   & 0.647                   & 0.628                   & -                                          \\
Aso      & 0.699                   & 0.714                   & -                                          \\
Hatoyama & 0.706                   & 0.710                   & \textless{}\textless{}\textless{}          \\
Kan      & 0.685                   & 0.669                   & \textgreater{}\textgreater{}\textgreater{} \\
Noda     & 0.697                   & 0.703                   & -                                          \\
Abe\_2   & 0.540                   & 0.626                   & \textless{}\textless{}\textless{}          \\
Suga     & 0.627                   & 0.711                   & \textless{}\textless{}\textless{}          \\ \hline
\end{tabular}
\caption{
The sentiments for the previous PMs before and after their change timing and the result of comparison.
The meaning of brackets is the same as Table~\ref{fig:common_vs_indiv}.
}
\label{fig:prev_new}
\end{table}

\section{Discussion and Conclusion}

\subsection{Main Findings}
In this study, we attempted to explore and characterize the honeymoon effect on social media by analyzing Japanese Twitter users' reactions to eight changes in the PM in Japan.
The results showed that in the RDD for the sentiment of the previous and new PMs before and after their change (RQ1), we confirmed a significant jump in sentiment in four out of eight times, with one timing when sentiment significantly declined and no significant results in the remaining three times.
The results suggest that there is an overall tendency to have the honeymoon effect on social media, although not in all cases.

In the analysis of sentiment by topics (RQ2), the sentiment was often significantly lower for PM-specific topics than for common topics to all PMs (4 out of 8 times).
Comparisons among PMs using topics common to all PMs showed that topics with positive sentiment differed by the PM, demonstrating the effectiveness of social media in understanding the relative position of each PM and the issues expected of each PM. 

Finally, the user-based analysis (RQ3) revealed that the majority of users were replaced between those who tweeted about the previous PM before the resignation and those who tweeted about the new PM after the inauguration.
Furthermore, the sentiments of these replaced user bases were significantly different in many cases (7 out of 8 times), indicating that the user groups which the administration should appeal to on social media may differ depending on whether it is immediately after the inauguration or when the approval rating is declining.
In addition, the sentiment of users who tweeted about the same PM after the inauguration and before the resignation was significantly different (5 out of 8 times), and in many cases, the sentiment was lower before the resignation  (4 out of 5 times), confirming the honeymoon effect in a different way. 

\subsection{Implications for practitioners}
The honeymoon effect was observed in social media at the PM change timings, although not always, suggesting that if the honeymoon effect emerges, social media could be used as a gauge of public response to policies that are implemented during the honeymoon period. 
For example, the pressure Suga put on academia soon after taking office received the lowest sentiment, which is claimed to have actually influenced the unnecessary drop in his approval rating~\cite{EDITORIA16:online}.
We confirmed that social media had captured this impact, and considering that the typical approval surveys are conducted on a monthly basis, monitoring social media may have enabled a quicker reaction to the drop in sentiment. 

When a country's leadership changes, the new PM must draw lessons from the cases of the previous leaders in managing the government, but each politician also has unique characteristics and expectations, making it challenging to apply the lessons.
It could be easier to communicate with the public if we could identify the features that are popular for each PM, as we have just confirmed.
For example, sentiment is particularly higher about cabinet appointments for Suga, while Kishida has positive sentiment on foreign affairs. It would be important to recognize the strengths of each PM and take them into account in their communication strategies. 

PMs have greater control over whom they interact with on social media than in traditional media.
In this regard, it would be important to determine whether users are continuously vocal or one-time users in terms of different change timings and PMs.
In fact, as we confirmed, users who only tweeted about a new PM, not a prev PM, tend to generate higher sentiment right after the inauguration of the new PM.
Therefore, an essential tactic for enhancing general attitude is likely to be how to involve these one-time posters in communication. 

\subsection{Limitation and future work}
\noindent\textbf{RDD setting and more elaborate model:} In conducting the RDD, we used sentiments toward the \textit{different} PMs before and after the threshold.
We made this decision in order to be consistent with traditional surveys on approval ratings, but we could not fully take into account the impact on the potential impact of individual PM on sentiment.
Modeling the potential impact on the PM's personal sentiment is challenging; for example, using sentiment prior to the PM's inauguration would result in a different nature of sentiment because a PM's name recognition surges only after becoming PM in many cases.
However, the result of this study showed that only Abe, who was appointed to PM twice, received anomaly sentiments; thus, we assumed the same model could basically be applied to figures who served as PM for the first time.
Nevertheless, there is still room for improvement in the modeling of the elaboration of the impact of each individual PM, including the change of administrative party (LDP and DPJ), for example.
In addition, we did not explicitly incorporate exogenous influencing factors into our model (e.g., the 2007 financial crisis). In comparing the previous and new PMs, the impact of external factors is likely to be small because the collection timing of sentiments for two PMs is close. However, external factors would affect comparisons between the change timings regarding the extent of the honeymoon effect.

\noindent\textbf{Consideration of user scale:} The second Abe administration lasted eight years, during which time the number of Twitter users increased dramatically, which might affect the behavior of sentiment at an aggregated level. 
In fact, for the two PMs after Abe (i.e., Suga and Kishida), the behavior of sentiment toward them was specifically similar, including RDD results; thus, it is reasonable to anticipate that the analysis of PMs will be continued in the future in order to confirm this tendency with a sufficiently large number of users.
On the other hand, in the early days of the service (e.g., 2007), Twitter was not as popular as it is today, and as this study showed, there were fewer users who mentioned older PMs; thus, the results were likely to be less robust about the older the PMs. 
Since the number of previous tweets cannot be increased, it becomes even more crucial to carry out comparable experiments in the future to verify the validity of the honeymoon effect on social media. 

\noindent\textbf{Other countries and domains:} The honeymoon effect in actual approval ratings have also been observed in other countries besides Japan; thus, the analysis of other countries (e.g., the United Kingdom, which has a similar parliamentary system) is also expected to be conducted.
Politics is one of the main topics on Twitterverse worldwide~\cite{huszar2022algorithmic}; thus, if democracy works and people's will is effectively reflected in the selection of the national leader, the same observation as this study would be anticipated even in other countries.
Also, in addition to politics, companies have also used executive resignations to recover from scandals and business downturns; thus, it would be interesting to analyze the effect of such actions, including the linkage with stock prices.

\noindent\textbf{Stance detection:} Sentiment analysis is used in this study to examine the honeymoon effect. An alternative approach would be stance detection.  Since stance and sentiment are known to differ from one another~\cite{aldayel2021stance}, undertaking stance analysis is also useful for gaining a more multifaceted understanding of the phenomena.

\noindent\textbf{The end of the honeymoon period:} We focused on the honeymoon effect and analyzed it at the timing of PM changes.
On the other hand, it would also be important to analyze the timing of the end of the honeymoon period, i.e., when the high approval ratings of PMs start to fall.

\noindent\textbf{Topic selection and classification:} In this study, we propose a data-processing flow to classify unique topics and common topics for PMs. However, this flow requires manual adjustment at the end, and it would be desirable to be able to efficiently extract unique/common topics using a more sophisticated method.

\noindent\textbf{Aspect-based sentiment analysis:} We measured the sentiments for topics by calculating the average sentiments of tweets of a topic. An alternative approach would be to use aspect-based sentiment analysis~\cite{pontiki2016semeval}, which trains the sentiments of texts with certain aspects. 

\noindent\textbf{Potential missing tweets:} We retrieved the tweets in November 2022 via the Twitter API. Thus, we could not obtain tweets that were deleted~\cite{bhattacharya2016characterizing}. 
The analysis of recent PM changes may be more robust, considering that older tweets are more likely to be erased and that the volume of tweets about more recent PMs is larger.

\subsection{Ethical Considerations}    
We pay the utmost attention to the privacy of individuals in this study. 
We did not include personal names except for PM or account names in our analysis. We examined public tweets only.  
When sharing our tweet data, we will publish only a list of tweet IDs according to Twitter's guidelines.

\section*{Appendices}
\begin{figure}[!htbp]
\centering
\includegraphics[width=0.95\linewidth]{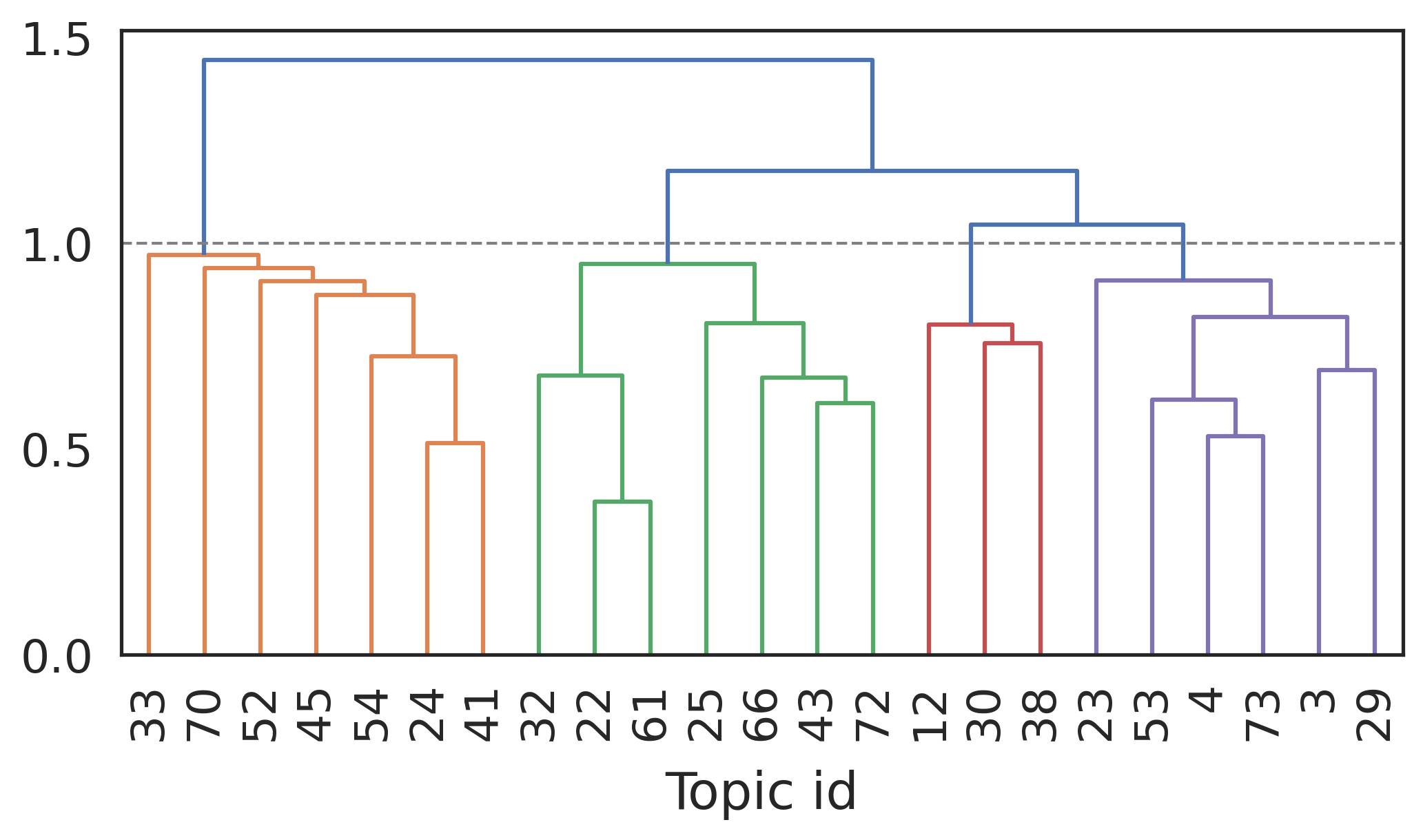}
\caption{
Dendrogram of hierarchical clustering of common topics. The distance between topics is the Hellinger distance for each topic's representative word calculated using the probability of belonging to all topics. The y-axis is the distance between clusters when they are combined using Ward's method. 
We set the threshold to 1 based on the shape of the dendrogram and the resulting cluster contents.
The x-axis contains the ids of each topic, the contents of which can be found in Table~\ref{table:common_topics}.
}
\label{fig:dendrogram}
\end{figure}

As mentioned in section~\ref{sec:topic_model}, we obtained 23 topics for eight new PMs, then we aggregated them using a hierarchical clustering method, the dendrogram of which is shown in Figure~\ref{fig:dendrogram}.
We manually determined the threshold for aggregation by looking at the dendrogram and also the contents of resulting clusters (Table~\ref{table:common_topics}).
As a result, we obtained four clusters, and we labeled them as \textit{Diplomacy}, \textit{Economy}, \textit{Cabinet personnel}, and \textit{Diet administration} by examining the representative words of the topics that comprise the cluster.

We found \textit{Diplomacy} as the topic on the United States, China, South Korea, and North Korea, with which Japan has close relations. 
For \textit{Economy}, we observed the topic of budgets, taxes, exchange rates, and wages. 
\textit{Cabinet personnel} include words for bureaucrats and ministers as well as some individual names, while \textit{Diet administration} consists of party names, approval ratings, press conferences, and speeches.
As for the tweet volume, \textit{Diplomacy} and \textit{Diet administration} are both the most popular topics, followed by \textit{Economy}, and finally \textit{Cabinet personnel}.

\begin{table*}[!htbp]
\centering
\scalebox{0.88}{
\begin{tabular}{lcp{10cm}rr}
\hline
Merged   topic                                                                         & Topic id & Top words                                                                                               & Tweet num & Total                \\ \hline
\multirow{7}{*}{Diplomacy}                                                             & 24       & Korea, Japan, Japan-Korea, President, summit meeting,   teleconference                                  & 111,824   & \multirow{7}{*}{505,899} \\
                                                                                       & 33       & Sign, North Korea, abduction, release, kidnapping                                                       & 50,269    &                          \\
                                                                                       & 41       & summit meeting, China, visit, diplomacy, minister                                                       & 90,324    &                          \\
                                                                                       & 45       & worship, Yasukuni shrine, Yasukuni, Yasukuni visit, China                                               & 38,771    &                          \\
                                                                                       & 52       & Okinawa, Futenma, relocation, governor, meeting                                                         & 59,881    &                          \\
                                                                                       & 54       & China, diplomacy, Senkaku, U.S., Biden                                                                  & 123,616   &                          \\
                                                                                       & 70       & speech, announcement, UN, reduction, summit                                                             & 31,214    &                          \\ \hline
\multirow{7}{*}{Economy}                                                               & 22       & bureaucrats, finance ministry, people, tax hike, Ozawa                                                  & 94,702    & \multirow{7}{*}{481,475} \\
                                                                                       & 25       & enterprise, wage increase, request, economy, professor                                                  & 30,518    &                          \\
                                                                                       & 32       & tax hike, consumption tax, income, finance, taxation                                                    & 67,136    &                          \\
                                                                                       & 43       & budget, government, direction, politics, meeting                                                        & 72,676    &                          \\
                                                                                       & 61       & people, policy, bureaucracy, tax hike, LDP                                                              & 68,554    &                          \\
                                                                                       & 66       & yen weakening, economy, policy, stock price, expectation                                                & 98,173    &                          \\
                                                                                       & 72       & economy, recovery, measures, employment, challenges                                                     & 49,716    &                          \\ \hline
\multirow{3}{*}{\begin{tabular}[c]{@{}l@{}}Cabinet \\      personnel\end{tabular}} & 12       & bureaucrats, personnel, appointment, LDP, charge                                                        & 46,961    & \multirow{3}{*}{170,082} \\
                                                                                       & 30       & inauguration, cabinet ministers, Tokyo, economy, press   conference                                     & 80,378    &                          \\
                                                                                       & 38       & minister, Renho, cabinet minister, minister in charge,   allegations                                    & 42,743    &                          \\ \hline
\multirow{6}{*}{\begin{tabular}[c]{@{}l@{}}Diet\\ administration\end{tabular}}   & 3        & policy speech, speech, concept, LDP, Osaka                                                              & 50,677    & \multirow{6}{*}{532,439} \\
                                                                                       & 4        & LDP, people, debate, opposition party, diet                                                             & 66,684    &                          \\
                                                                                       & 23       & approval rating, support, poll, polls, survey, cabinet   approval rating                                & 146,087   &                          \\
                                                                                       & 29       & Diet, opposition parties, questions from representatives,   policy speeches, extraordinary diet session & 115,450   &                          \\
                                                                                       & 53       & question, press, pressure, incident, problem                                                            & 43,924    &                          \\
                                                                                       & 73       & people, language, politics, society, unhappiness                                                        & 109,617   &                          \\ \hline
\end{tabular}
}
\caption{
Common topics that are aggregated by clustering.
``Merged topic'' indicates the name of the cluster as labeled by the authors.
``Topic id'' is a random id assigned to a topic by BTM. 
``Top words'' are the five most possible words belonging to each topic.
The number of tweets for each topic and cluster is also shown.
}
\label{table:common_topics}
\end{table*}

\bibliographystyle{ACM-Reference-Format}
\bibliography{main}


\end{document}